# Enhancement of visible-light-induced photocurrent and photocatalytic activity of V and N codoped TiO$_2$ nanotube array films

Min Zhang[*], Dandan Lu, Zhihua Zhang, Jianjun Yang[*]

*Key Laboratory for Special Functional Materials of Ministry of Education, Henan University, Kaifeng 475004, PR China*

**Email:** zm1012@henu.edu.cn, yangjianjun@henu.edu.cn

Tel: (86)371- 23881358, Fax: (86)371- 23881358

**Abstract:** Highly ordered TiO$_2$ nanotube arrays (TNAs) codoped with V and N were synthesized by electrochemical anodization in association with hydrothermal treatment. The samples were characterized by field emission scanning electron microscopy, X-ray diffraction and X-ray photoelectron spectroscopy. The photocurrent and photocatalytic activity of codoped TiO$_2$ nanotube arrays were investigated under visible light irradiation. Moreover, the production of hydroxyl radicals (·OH) on the surface of visible light-irradiated samples is detected by a photoluminescence technique using terephthalic acid (TA) as a probe molecule. It was found that the V+N co-doped TiO$_2$ nanotube arrays showed remarkably enhanced photocurrent and photocatalytic activity than undoped sample due to the V and N codoping.

**Keywords**：

TiO$_2$；phtocatalytic activity； nanotube； photocurrent

**Introduction**

Highly ordered TiO$_2$ nanotube arrays (TNAs) prepared by anodization of titanium have attracted much attention in recent years, because of their large surface area, high photocatalytic activity, vectorial charge transfer and good recyclability as well as unique nanostructure and promising application in solar light conversion as compared with conventional TiO$_2$ powder [1]. However, the use of nanostructured TiO$_2$ materials as a visible light photocatalyst is still limited, since their large bandgap of 3.2 eV can only be activated by ultraviolet radiation that constitutes only a small fraction (3%~5%) of the solar spectrum. On the other hand, the recombination rate of photogenerated electron−hole pairs is too high to be used for organic pollutant degradation in practice [2].

To surmount these problems, doping of TiO$_2$, especially noncompensated cation-anion codoping, had been used to increase availability of visible light and improve the separation efficiency of photo-induced electrons and holes. The coupling of one dopant with the second has been proposed to enable a reduction in the number of carrier recombination centers by proposed charge equilibrium mechanism as well as enhancing visible light absorbance by increasing the solubility limit of dopants. It is generally accepted that electronic coupling between two dopants is the key to realize synergistic effects of codoping on photocatalytic activity [3]. For example, Yin et al. and Long et al. separately proposed that the injection of 3*d* or 4*d* transition metals (such as Mo, W and V) can increase the solubility limits of N and the band edges of TiO$_2$ can be modified by co-dopants to significantly shift the valence band edge up based on density-functional theory calculations [4, 5].

Among the current reports of single ion doping into anatase TiO$_2$, N-doping and V-doping are noteworthy. Vanadium is a transition metal with multiple characteristics which improve TiO$_2$ absorptivity. For example, V-doped TiO$_2$ with different valance state of the V and Ti (V$^{0/1+/2+/3+/4+/5+}$ and Ti$^{1+/2+/3+/4+}$) exhibits a difference in oxidation activity [6]. V doping was responsible for increases of superficial hydroxyl groups and electron transfer, resulting in faster interactions between the adsorbate and adsorbent [7]. The ionic radius of V is close to that of Ti, allowing it to be easily doped into TiO$_2$ [8, 9].

So, vanadium ion doping, as one of the best alternatives to enhance the separation efficiency of electron-hole pairs and the visible light-driven photoactivity, has attracted extensive interest.

N doping is also considered an effective approach to improve visible spectral absorption and photocatalytic activity [10, 11]. So the codoping of V and N into the lattice of $TiO_2$ is of particular significance. For example, V and N ions were reported to incorporate into anatase lattice of $TiO_2$ samples. And these co-doped $TiO_2$ photocatalysts showed enhanced photocatalytic activities for the degradation of methylene blue compared with monodoped $TiO_2$ [12]. Wang et al. synthesized V-N codoped $TiO_2$ nanocatalysts using a novel two-phase hydrothermal method applied in hazardous PCP-Na decomposition [13]. Recently, Zou et al. investigated the energetic and electronic properties of N-V-codoped anatase $TiO_2$ (101) by first principles calculations to elucidate the relationship between the electronic structure and the photocatalytic performance of N-V-codoped $TiO_2$ [14]. They also investigated the adsorption and decomposition behaviors of water molecule on the surface of N-V codoped anatase $TiO_2$ (101). It is found that the N−V-codoping could broaden the absorption spectrum of anatase $TiO_2$ to the visible-light region, and could enhance its quantum efficiency [15].

Nevertheless, few reports are currently available about V and N co-doping of $TiO_2$ nanotube arrays and their photoelectrochemical and photocatalytic properties as well. In this work, we adopt a sequence of electrochemical anodization and hydrothermal synthesis technique to fabricate V and N co-doped $TiO_2$ nanotube array films. For V and N codoped TNAs, the influence of morphology, structure and doping concentration on photoelectrochemical performance as well as photocatalytic activity for the degradation of methylene blue (denoted as MB) were investigated under visible light irradiation (≥420 nm).

**Experimental**

V, N co-doped $TiO_2$ nanotube arrays (TNAs) were fabricated by a combination of electrochemical anodization and a hydrothermal reaction. Firstly, highly ordered

TNAs were fabricated on a Ti substrate in a mixed electrolyte solution of ethylene glycol containing $NH_4F$ and deionized water by a two-step electrochemical anodic oxidation process according to our previous reports [16]. Then the amorphous TNAs were annealed at 500 °C for 3 h with a heating rate of 10 °C/min in air ambience to obtain crystalline phase. Interstitial nitrogen species were formed in the TNAs due to the electrolyte containing $NH_4F$ or a lesser extent atmospheric nitrogen dissolved into the electrolyte during anodization [17]. We denoted these single N-doped TNAs samples as $N-TiO_2$.

V and N codoped TNAs were prepared by a hydrothermal process. As-prepared $N-TiO_2$ was immersed in a Teflon-lined autoclave (120 mL, Parr Instrument) containing approximately 60 mL of $NH_4VO_3$ solution (with different concentration 0.5, 1, 3 and 5 wt %) as the source of both V and N. The samples were hydrothermally heated at 180 °C for 5 h and then naturally cooled down to room temperature. Finally, all as-prepared samples were rinsed with deionized water and dried under high purity $N_2$ stream. The corresponding products (0.5%, 1%, 3%, 5%) were labeled as VN0.5, VN1, VN3, VN5.

The morphologies of the obtained samples were characterized with field emission scanning electron microscope (FESEM, FEI Nova NanoSEM 450). Phase structures of photocatalysts were analyzed by X-ray diffraction (XRD, Philips X'Pert Pro X-ray diffractometer; Cu Kα radiation, $\lambda$ =0.15418 nm). Surface composition of nanotube arrays was analyzed by X-ray photoelectron spectroscope (XPS) on a Kratos Axis Ultra System with monochromatic Al Kα X-rays (1486.6 eV). The binding energies (BE) were normalized to the signal for surface adventitious carbon at 284.8 eV.

Photoelectrochemical experiment was monitored by an electrochemical workstation (IM6ex, Germany). TNAs film (an active area of 4 $cm^2$) and platinum foil electrode were used as working electrode and counter electrode, and saturated calomel electrode (SCE) acted as reference electrode, respectively. 1 M KOH aqueous solution was used as a supporting electrolyte and purged with $N_2$ for 20 min before measurement to remove dissolved oxygen. A 300 W xenon lamp (PLS-SXE300,

Beijing Bofeilai Technology Co, Ltd.) with a cutoff filter ($\lambda \geqslant 420$ nm) was used as the visible light source. The intensity of visible light irradiated on tested samples was 16 mW·cm$^{-2}$.

MB solution was chosen as a target pollutant, with an initial concentration of 10mg/L, since it is non-biodegradable and extensively used in textile industry. TNAs films (active area of 4 cm$^2$) were dipped into 25 mL of MB solution. A 300W xenon lamp with a cutoff filter ($\lambda \geqslant 420$ nm) as visible light source in the photocatalytic experiments. Prior to irradiation, the solution with TiO$_2$ sample was magnetically stirred in the dark for 30 min to establish adsorption–desorption equilibrium. The changes of concentration were monitored for analyzing the photocatalytic activity at 30 min irradiation time intervals by measuring the absorption at 664 nm using an SP-2000 spectrophotometer. Though the amount of MB adsorption is different for different samples before light irradiaiton, the adsorption amount has been subtracted when the photocatalytic activity is calculated. The photocatalytic activity is calculated by $(C_0 - C)/C_0 \times 100\%$, $C_o$ is MB concentration after adsorption–desorption equilibrium in dark, C is MB concentration after photocatalytic reaction under visible light irradiation. This calculation method of photocatalytic activity can exclude the influence of adsorption amount due to the amount of MB adsorption is different for different samples.

Terephthalic acid (denoted as TA) was used as a probe to monitor the change of hydroxyl radical content of all TNAs samples under visible light irradiation, since TA can readily react with •OH radical to generate TAOH which emits fluorescence around 426 nm under the excitation of its own absorption band at 312 nm. Experimental procedures are similar to the measurement of photocatalytic activity except that MB aqueous solution was replaced by TA aqueous solution (6 mmol·L$^{-1}$) with a concentration of NaOH solution (20 mmol·L$^{-1}$). Photoluminescence (PL) spectra of generated TAOH were measured on a Gilden FluoroSENS fluorescence spectrometer. The content of •OH radical over different TNAs surface under visible light irradiation was compared based on PL data.

## Results and discussion

Top-view FESEM images of N-TiO$_2$ and VN3 samples are presented in Fig. 1. Before codoping with vanadium via hydrothermal process, the nanotubes of N-TiO$_2$ sample are open at the top end and have an average diameter of 110 nm ( Fig. 1a), and their nanotube array structure can be clearly observed from the inset of Fig.1a. Moreover, SEM observation indicates that there is no apparent structural transformation of the TNAs sample after hydrothermal co-doping process (Fig.1b). The appropriate doping amount of V and N does not change the morphology of nanotubes. Fig. 1c shows the EDX data of codoped VN3 sample. There are N and V peaks in the EDX pattern that gave convincing evidence for the presence of nitrogen and vanadium on the surface of TiO$_2$ nanotube arrays. The mole ratio of N/Ti and V/Ti was estimated to be ca.12.7% and 1.8%, respectively.

Fig. 2 shows the XRD patterns of the N-TiO$_2$ and V, N codoped TNAs with various doping amounts. The diffraction peaks of all samples were ascribed to pure anatase TiO$_2$ diffraction pattern, which were consistent with the values in the standard card (JCPDS 21-1272) [18]. The results indicate that V and N codoping has no effect on the crystal structure and phase composition of TNAs. Moreover, no V$_2$O$_5$ phase could be observed in all the XRD patterns. Accordingly, we propose that V ions may be incorporated into the titania lattice and replaced titanium ions to form V-O-Ti bonds.

Transition metal ions as dopant incorporated into other compounds might distort the crystal lattice of the doped materials. Thus, we enlarged the anatase (101) plane of samples in the inset of Fig. 2b to check the structural change. Compared with N-TiO$_2$, peak position of V, N codoped TiO$_2$ shifted toward a higher diffraction angle gradually, suggesting that V ions might be successfully incorporated into the crystal lattice of anatase TiO$_2$ as vanadyl groups (V$^{4+}$) or polymeric vanadates (V$^{5+}$) and substituted for Ti$^{4+}$, because the ionic radii of V$^{4+}$(0.72 Å) and V$^{5+}$(0.68 Å) were both slightly smaller than that of Ti$^{4+}$(0.75 Å) [19].

Figure 3 presents the high-resolution XPS spectra of V 2p, N 1s, Ti 2p and O 1s regions for N-TiO$_2$ and VN3 samples. Fig. 3a shows the high resolution XPS spectra and corresponding fitted XPS for the N 1s region of N-TiO$_2$ and VN3. A broad peak extending from 397 eV to 403 eV is observed for all samples. The center of the N1s peak locates at ca. 399.7 eV and 399.4 eV for N-TiO$_2$ and VN3 sample, respectively. These two peaks are much higher than that of typical binding energy of N1s (396.9 eV) in TiN [20], indicating that the N atoms in all samples interact strongly with O atoms [21]. The binding energy of 399.7 eV and 399.4 eV are attributed to oxidized nitrogen similar to NO$_x$ species, meaning Ti-N-O linkage possibly formed on the surface of N-TiO$_2$ and VN3 samples [21, 22].

Fig. 3b presents the high resolution V 2p spectra of VN3 sample. The V 2p peaks appearing at about 516.3 eV and 523.8 eV are assigned to 2p$_{3/2}$ and 2p$_{1/2}$ electronic states of V$^{4+}$, 516.9 eV and 524.4 eV are attributed to 2p$_{3/2}$ and 2p$_{1/2}$ of V$^{5+}$, respectively [23, 24]. This indicates that the V$^{4+}$ and V$^{5+}$ions were successfully incorporated into the crystal lattice of anatase TiO$_2$ and substituted for Ti$^{4+}$ions, which is accordance with relevant XRD analytical results. The concentrations of V and N in VN3 derived from XPS analysis were 3.38% and 4.21% (at.%), respectively. The molar ratios of N/Ti on the surface of N-TiO$_2$ and VN3 were 2.89% and 14.04%, indicating obvious increase of N doping content by hydrothermal codoping with V and N.

Fig.3c and d present high-resolution spectra of Ti 2p and O1s of N-TiO2 and VN3 samples. Compared with N-TiO$_2$ sample, a significant negative shift is observed for Ti 2p in Fig.3c and O 1s in Fig.3d after V and N were doped into TiO$_2$ (VN3) by hydrothermal codoping process. One possible explanation is that Ti$^{4+}$ has been reduced to Ti$^{3+}$, which results in a negative shift of binding energies of Ti 2p for the V, N codoped TiO$_2$ samples. On the other hand, oxygen molecules can be dissociatively absorbed on the oxygen vacancies induced by doping N, thereby leading to a slight shift to lower binding energy of O 1s of TiO$_2$ lattice oxygen (Ti−O−Ti) [25].

To investigate the effect of V, N codoping of TiO$_2$ nanotube arrays on charge carriers separation and electron transfer process, a series of the PEC experiments were conducted. Fig.4a shows the photocurrent-potential dependence of N-TiO$_2$ and V,N codoped TNAs samples under visible light irradiation. The photocurrent density of

V,N codoped TNAs is significantly high than that of N-TiO$_2$ at all bias potential, indicating a lower recombination of photogenerated electrons and holes after V and N codoping. Moreover, enhanced photocurrent density was observed with the increase of vanadium doping amount and a highest photocurrent value of 5.8 µA/cm$^2$ was obtained for VN3 sample. However, further increasing the doping amount, the photocurrent decreased to 1.2 µA/cm$^2$ for VN5. It indicated that there is an optimal concentration for V,N codoping.

The generated photocurrents also increased significantly with an increase in the applied voltage. The higher the applied potential, the higher the photocurrent generated by all samples. The small observed photocurrent of N-TiO$_2$ sample could be ascribed to possible pollutants on electrode surfaces or impurities or species in the KOH solution. The high photocurrent conversion efficiency of V,N co-doped TiO$_2$ nanotube array films may be attributed to improved optical absorption and increased in the crystallinity caused by the hydrothermal co-doping process.

Furthermore, the transient photocurrents measured at fixed bias voltage of 0.8 V vs. SCE with a visible light pulse of 50s were also shown in Fig.4b. All samples showed good photoresponses and highly reproducible for numerous on–off cycles under the light on and light off conditions. V, N codoped TNAs exhibited higher photocurrents than that of N-TiO$_2$ under visible light irradiation. The VN3 sample exhibited highest photocurrent of about 4.5 µA/cm$^2$. These results further inferred that V, N codoped TiO$_2$ nanotube arrays possess good photo-responsive to generate and separate photo-induced electrons and holes [26]. However, excessive vanadium content may cause a detrimental effect, which acted as recombination centers to trap the charge carriers and resulted in a low quantum yield. From above PEC experimental results optimum content of V codoped into TiO$_2$ nanotube arrays play an important role in maximizing the photocurrent density mainly attributed to the effective charge carriers separation and improve the charge carriers transportation.

Photocatalytic activities of all samples were evaluated by using methylene blue as the target pollutant. Fig.5a shows the degradation curves of MB over N-TiO$_2$ and V, N codoped TNAs under the visible light. It reveals that the order of photocatalytic activity of TNAs was VN3 > VN1 > VN0.5 > N-TiO$_2$ > VN5. The photocatalytic activity of V, N codoped TNAs was enhanced and then decreased with an increase in the doping content of vanadium. V, N codoped TNAs possess much higher photocatalytic activity under visible light irradiation than N-TiO$_2$ sample except VN5 sample. The overloaded vanadium dopant of VN5 may result in channel plugging and fast recombination of hole and electron pairs [27]. Fig.5b shows the kinetic linear simulation curve of MB over N-TiO$_2$ and V, N codoped TNAs under the visible light. The decomposition kinetics was analyzed according to physical chemistry principles. The above degradation reactions followed a Langmuir–Hinshelwood apparent first order kinetics model due to the low initial concentrations of reactants. The relative concentrations of MB were fitted by the apparent first order rate equation as follows: $-\ln(C/C_0) = kt$, Where k is the apparent reaction rate constant. VN3 sample showed the highest lowing rate in the degradation of MB with a rate constant of 0.0136 min$^{-1}$, which was approximately 3.60 times higher than that of N-TiO$_2$.

The separation efficiencies of photo-induced electrons and holes under visible light irradiation of N-TiO$_2$ and V, N codoped TNAs samples were further studied by detecting the production of hydroxyl radicals. Hydroxyl radicals are mainly generated from the oxidative reaction between photo-induced holes and H$_2$O or OH$^-$, and the reductive reaction between photo-induced electrons, O$_2$ and H$^+$. So that the amount of hydroxyl radicals produced corresponds to the separation efficiencies of photo-induced electrons and holes as well as the photocatalytic activity. Moreover, the experiment of measuring active hydroxyl radicals can avoid the dye self-photosensitization effect and prove photocatalytic activity under visible light. Therefore, to further confirm the difference of photocatalytic activity of the prepared samples, the hydroxyl radical was monitored by a PL technique using TA as the fluorescence probe on the surface of TNAs under visible light illumination.

Fig.6a shows the changes of PL spectra of TA solution with irradiation time for VN3 sample. A gradual increase in PL intensity at about 425 nm is observed with increasing irradiation time for VN3 sample. However, no PL increase was observed in the absence of visible light or TNAs. This suggests that the fluorescence is originated from the chemical reactions between TA and •OH produced at the codoped TNAs /water interface under visible light irradiation [28].

Fig.6b shows the comparison of PL intensity of TA solution for N-TiO$_2$ and V+N codoped TNAs samples under visible light irradiation for 1h. The amount of hydroxyl radicals produced over VN3 is higher than those over all other samples, which is consistent with the photocatalytic activity result shown in Fig.5. Hydroxyl radical experiments further confirm that hydroxyl radicals are active species during photocatalytic reactions. On the other hand, the formation rate of hydroxyl radicals has a positive relation with the photocatalytic activity [29].

## Conclusions

In summary, visible light responsive TiO$_2$ nanotube arrays by V and N codoping are successfully fabricated by a simple two-step method. All V and N co-doped TNAs exhibit fine tubular structures after hydrothermal codoping process. XPS data indicate that N is present in the forms of Ti-N-O and V incorporates into the lattice of TiO$_2$ in V+N co-doped TNAs. V and N doping amount exhibit a great influence on the photocatalytic activity and photoelectrochemical properties of the TiO$_2$ nanotube array films. VN3 sample presents highest photocatalytic activity for MB degradation, largest formation rate of hydroxyl radicals and photocurrent under visible light irradiation due to the synergetic effects of several factors including effective separation of electron-hole pairs and remaining tubular structures. This study provides a new approach to successfully enhance visible light photocatalytic activity.

## Acknowledgements

The authors thank the National Natural Science Foundation of China (no.21203054) and Program for Changjiang Scholars and Innovation Research Team in University (no. PCS IRT1126).

**Figure Captions**

Figure 1. FESEM top views for N-TiO$_2$ (a) and VN3 (b) samples. EDX of VN3 sample (c).

Figure 2. XRD patterns (a) and the enlargement of the anatase (101) peaks (b) for N-TiO$_2$, VN0.5, VN1, VN3 and VN5 samples.

Figure 3. High-resolution XPS spectra of (a) N 1s, (b) V 2p, (c) Ti 2p and (d) O1s for N-TiO$_2$ and VN3 samples.

Figure 4. Photocurrent density vs applied potential curves (a) and photocurrent responses in light on-off process at applied voltage of 0.8 V (vs SCE) (b) under visible light irradiation for N-TiO$_2$, VN0.5, VN1, VN3 and VN5.

Figure 5. Concentration change (a) and photocatalytic degradation rate (b) of MB in the presence of N-TiO$_2$ and V+N codoped TiO$_2$ nanotube arrays under visible light irradiation.

Figure 6. (a) PL spectral changes with visible light irradiation time on VN3 sample in TA solution. (b) Comparison of PL spectra of the TiO$_2$ nanotube arrays before and after codoping in TA solution under visible light irradiation at a fixed 1h.

**Figure 1**

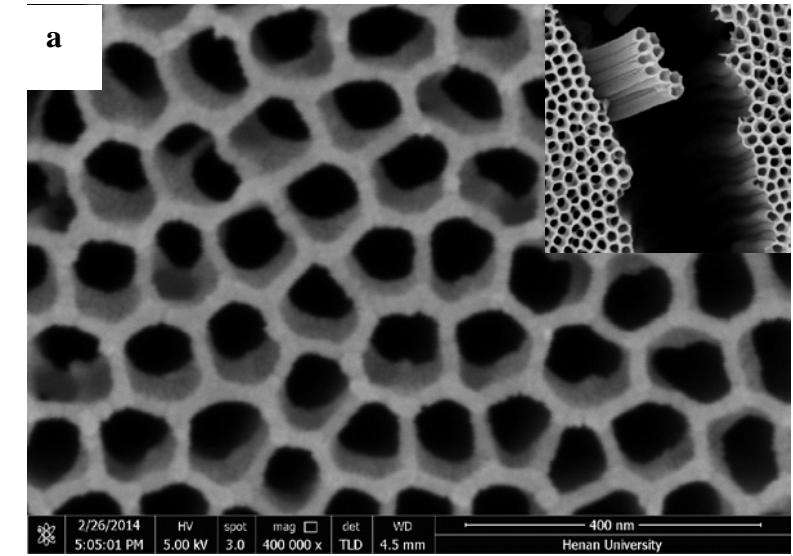

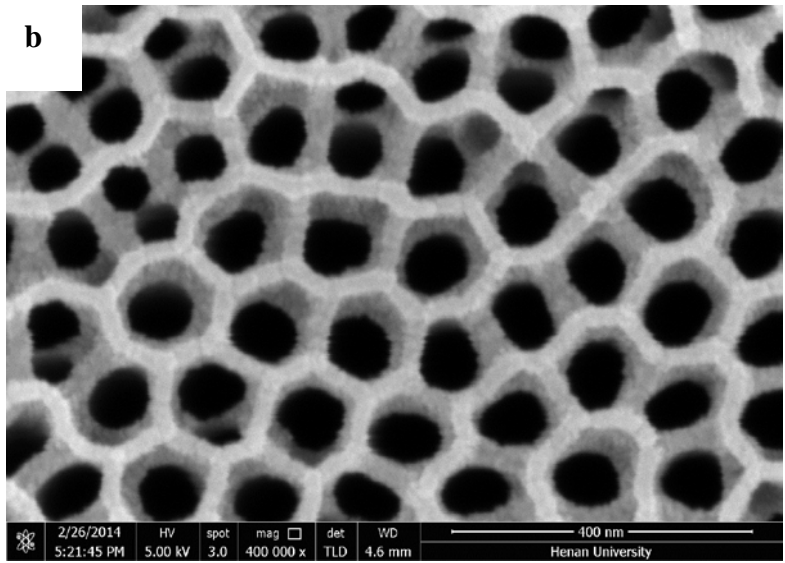

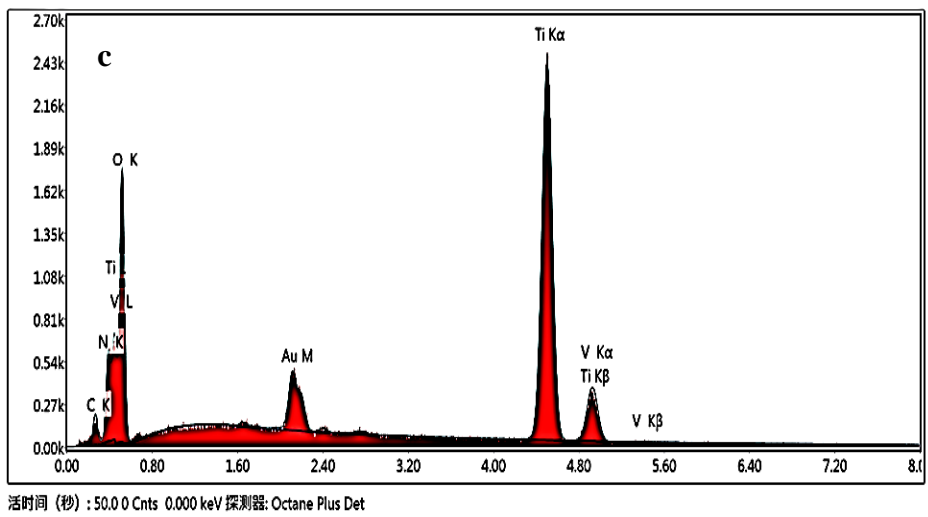

Figure 1. FESEM top views for N-TiO$_2$ (a) and VN3 (b) samples. EDX of VN3 sample (c).

**Figure 2**

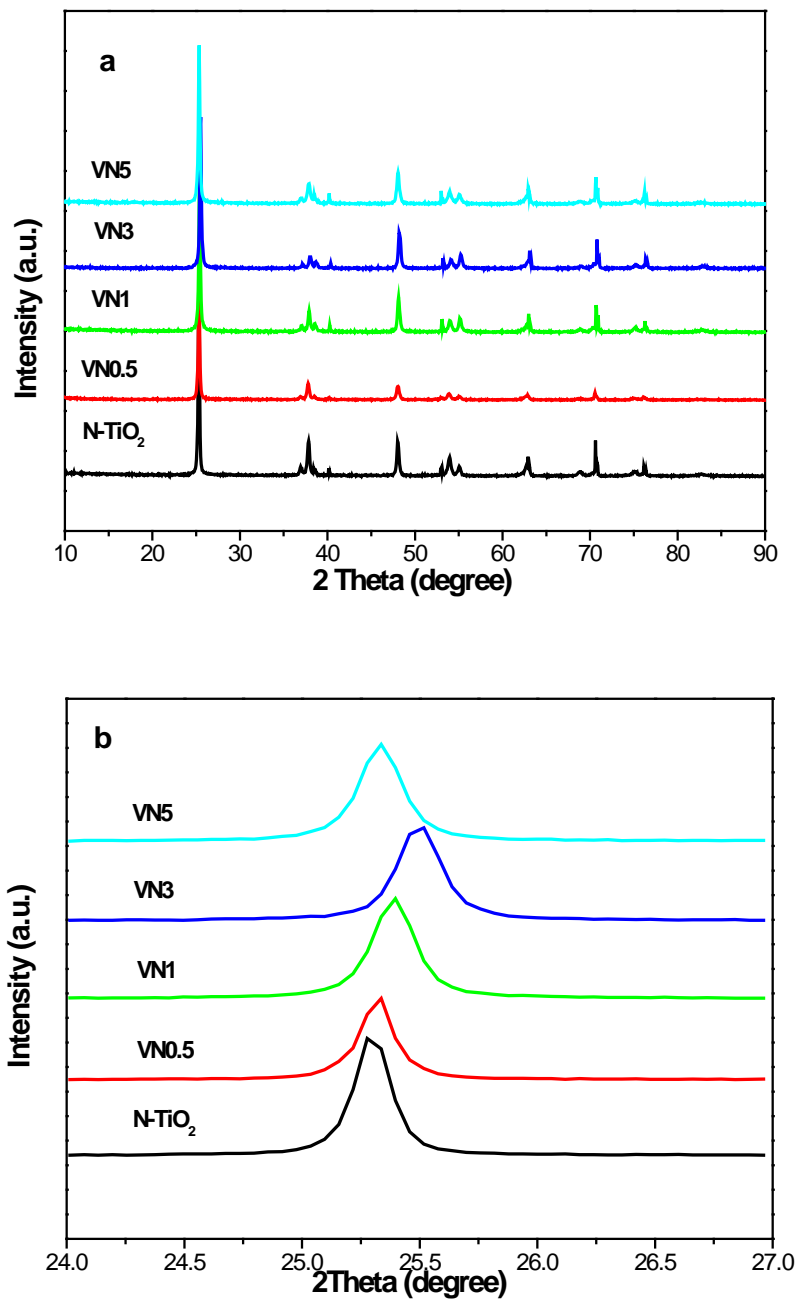

Figure 2. XRD patterns (a) and the enlargement of the anatase (101) peaks (b) for N-TiO$_2$, VN0.5, VN1, VN3 and VN5 samples.

**Figure 3**

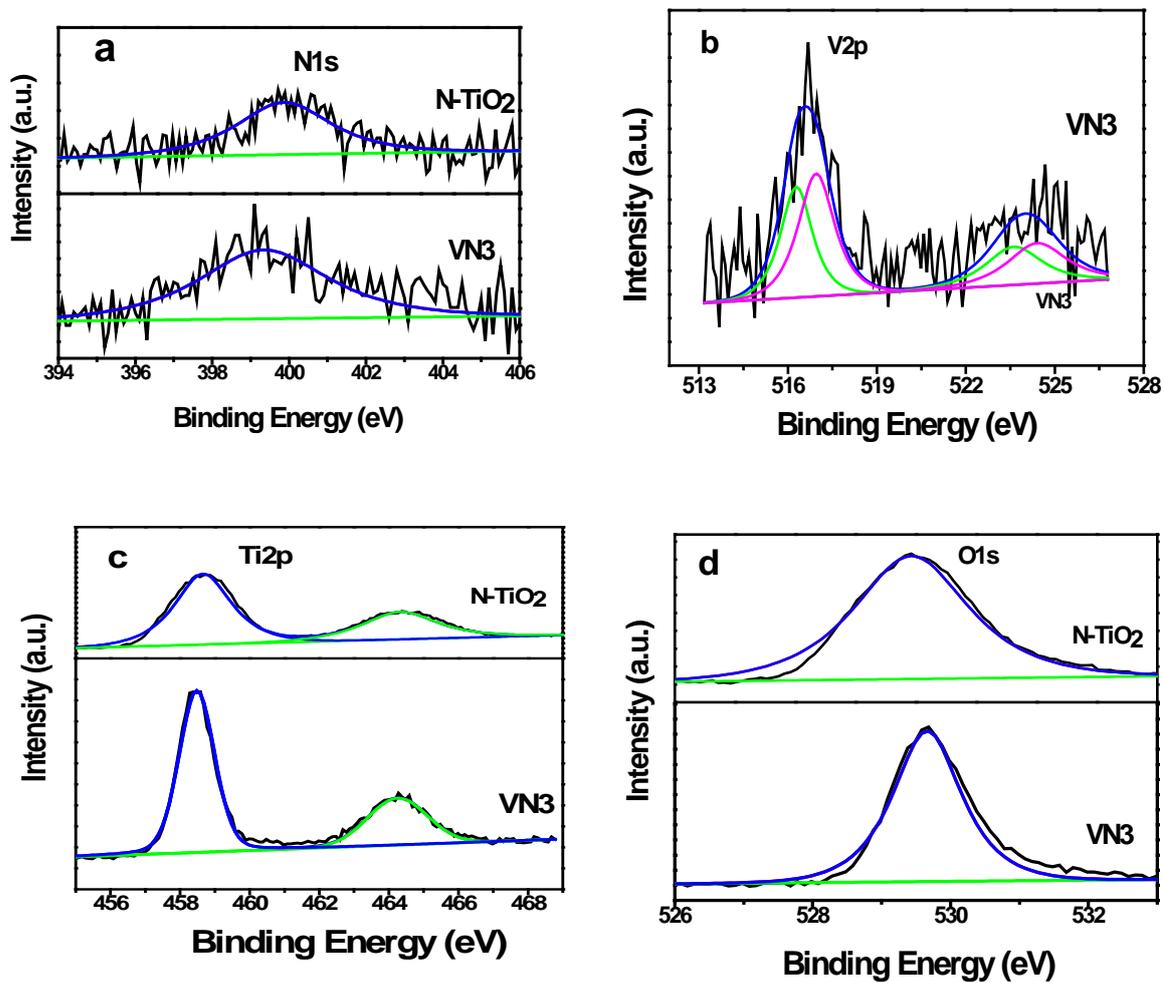

Figure 3. High-resolution XPS spectra of (a) N 1s, (b) V 2p, (c) Ti 2p and (d) O1s for N-TiO$_2$ and VN3 samples.

**Figure 4**

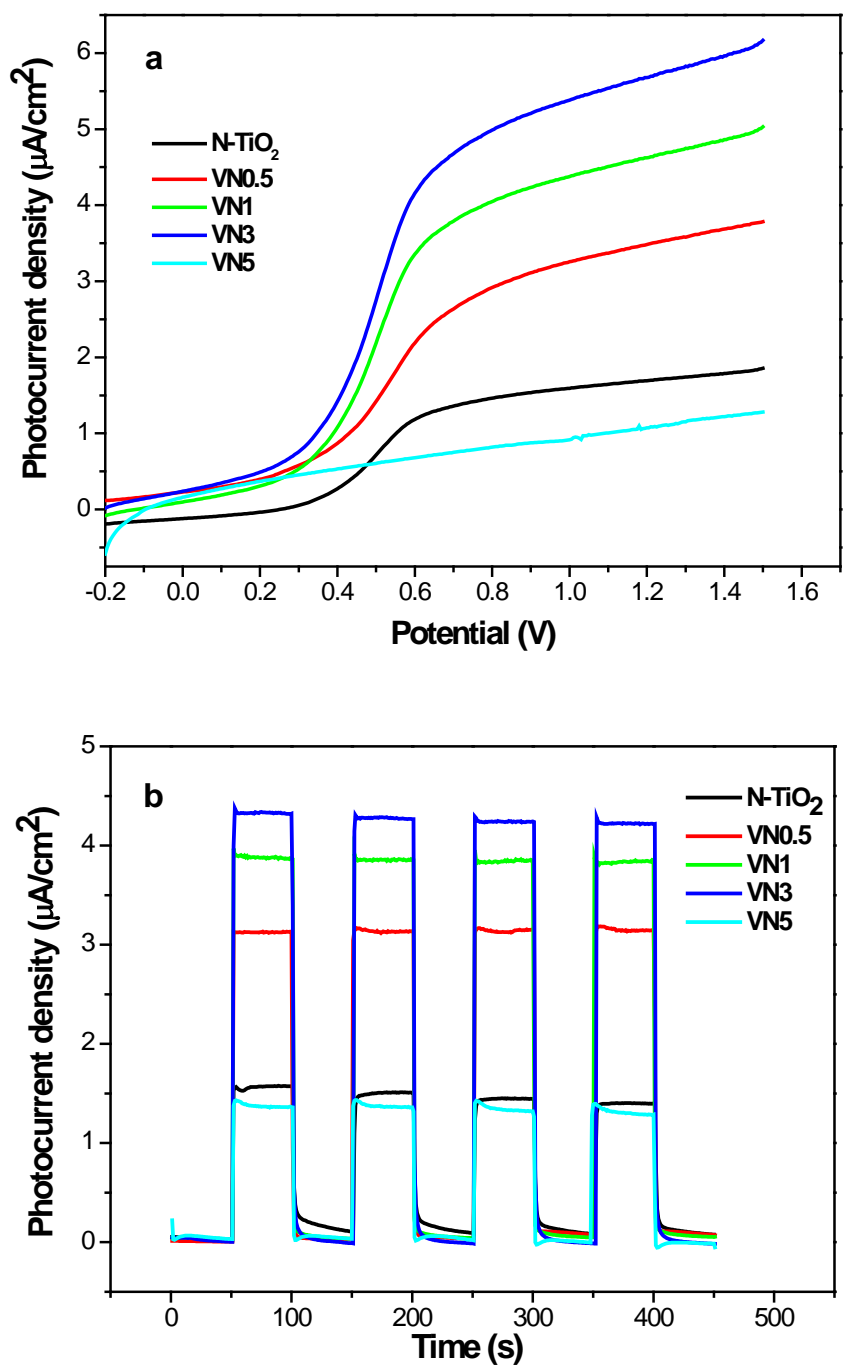

Figure 4. Photocurrent density vs applied potential curves (a) and photocurrent responses in light on-off process at applied voltage of 0.8 V (vs SCE) (b) under visible light irradiation for N-TiO$_2$, VN0.5, VN1, VN3 and VN5.

**Figure 5**

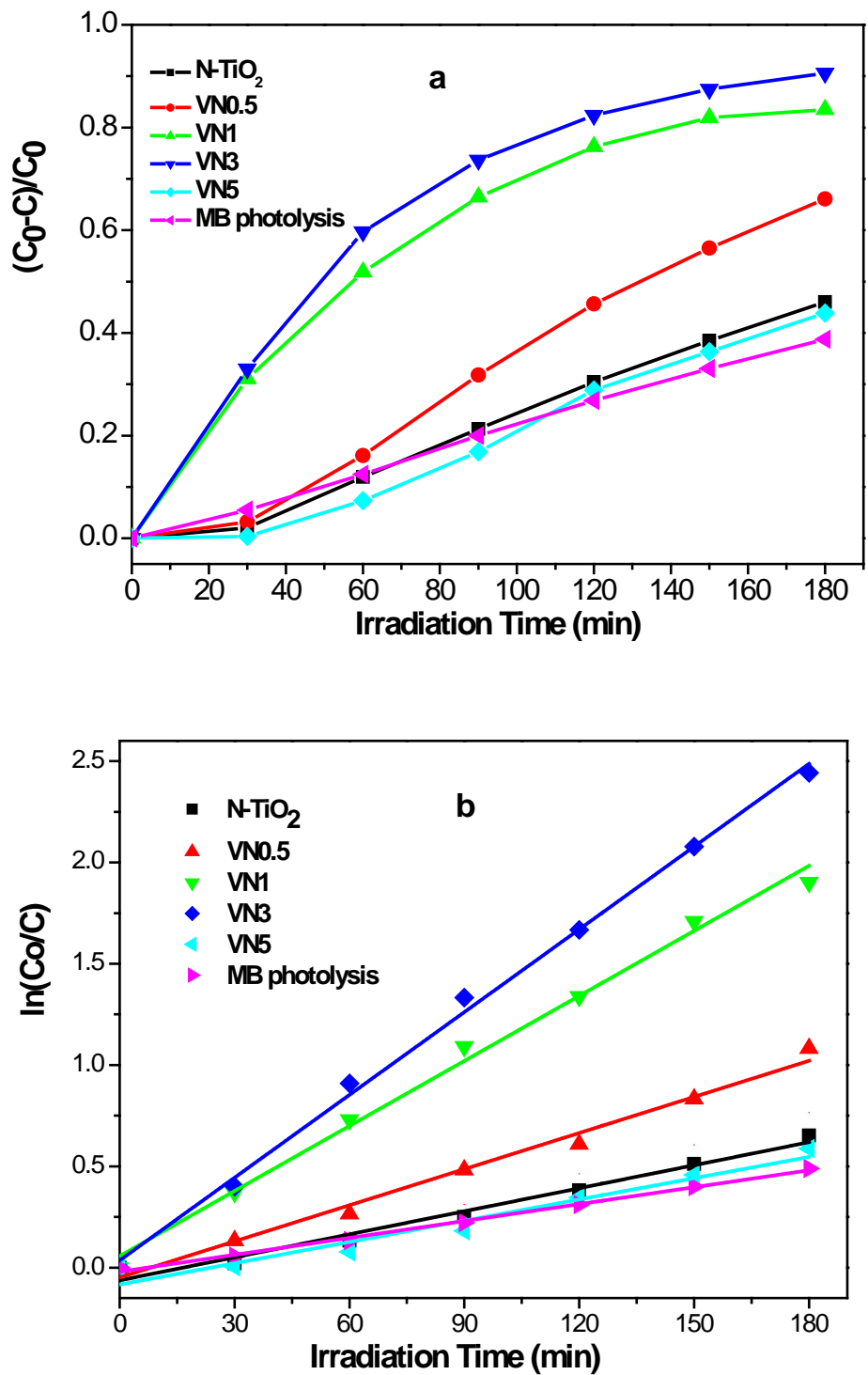

Figure 5. Concentration change (a) and photocatalytic degradation rate (b) of MB in the presence of N-TiO$_2$ and V+N codoped TiO$_2$ nanotube arrays under visible light irradiation.

**Figure 6**

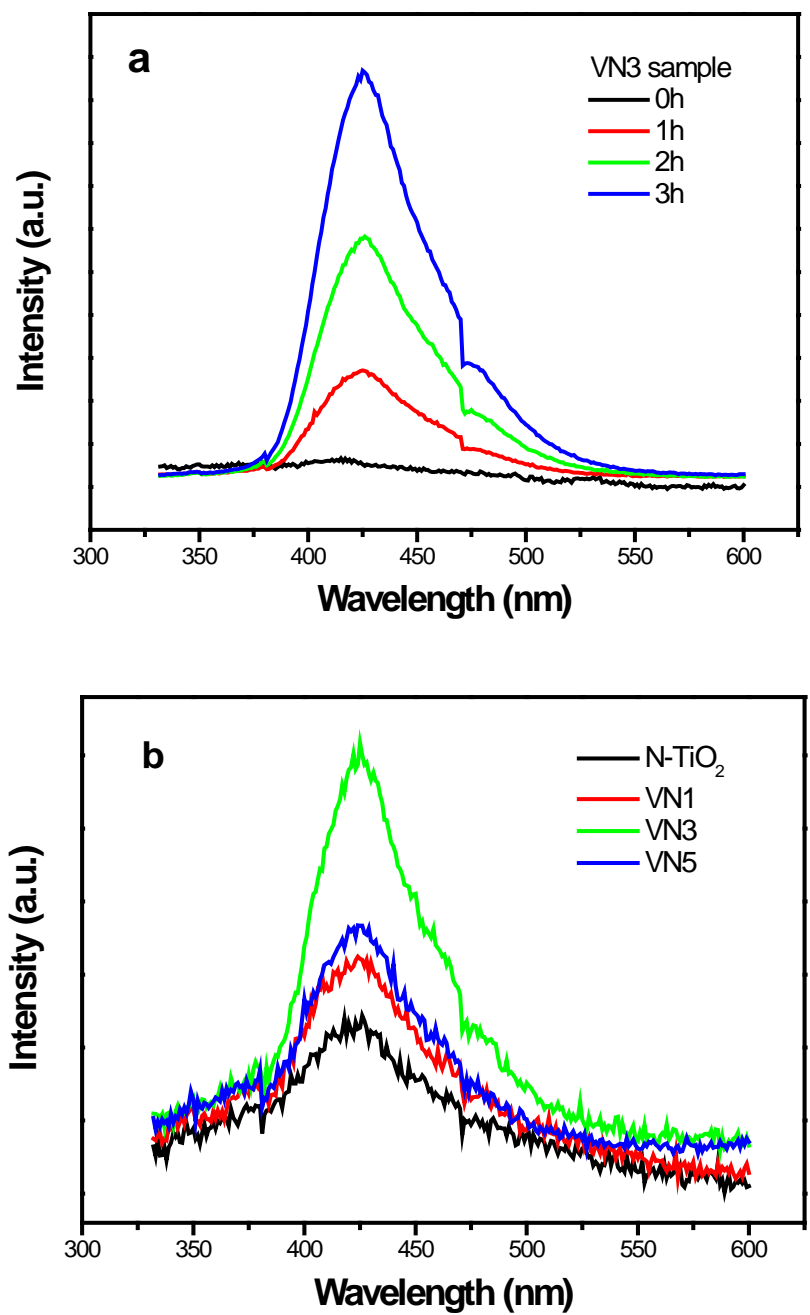

Figure 6. (a) PL spectral changes with visible light irradiation time on VN3 sample in TA solution. (b) Comparison of PL spectra of the $TiO_2$ nanotube arrays before and after codoping in TA solution under visible light irradiation at a fixed 1h.